# New evaluation tool for predicting disability pension risk among Finnish public sector employees

Petra Sohlman
Keva, Helsinki, Finland
ORCID: 0009-0001-2649-0736
petra.sohlman@keva.fi

Risto Louhi
Keva, Helsinki, Finland
ORCID: 0009-0003-1602-0962
risto.louhi@keva.fi

Janne Salonen
Keva, Helsinki, Finland, Tampere University, Tampere, Finland
ORCID: 0000-0002-0595-6226
janne.salonen@keva.fi

**Abstract** Using unique research data, we investigate disability retirement risk under the statutory public sector pension scheme in Finland. The statistical analysis yields two indicators: risk for upcoming permanent disability pension and critical duration of sickness absence days for public sector occupations. Statistical analysis is based on logistic regression model where the outcome is the disability pension, using sickness benefit spells and other individual background information as covariates. The results underline the importance of minimizing the sickness spells and their duration to the risk and reveal differences in risk across occupations. We conclude that the proposed risk model is a promising tool which can help employers and the pension industry in preventing permanent disability.

**Keywords** design thinking; logistic regression; occupation; sickness absence; unit-level data

## 1. Introduction

Despite the decrease over the last decade in the amount of new disability pension retirees in many countries, multidimensional challenges remain because of fragmentary working lives. In statutory pension systems the costs of pensions fall on pension providers and employers along with employees. There is a wide range of consequences of disability, in particular of disability pensions (DP), beyond the pension industry because of their significant effect on the employers and employees with challenges in work ability.

Work ability often diminishes over time, offering a time frame for preventive interventions. Sickness absence (SA) spells often precede permanent disability retirement, which is why so much attention is paid to early interventions, for example, by occupational health care provided by the employers. In addition, the statutory pension systems, including national pension systems, hold various schemes for handling sickness and disability benefits for the working age population.

In the Finnish context there exists a body of literature on the factors affecting disability retirement and SAs. For example, the labor market outcomes after sickness spells have recently been studied from different perspectives (e.g., Leino-Arjas et al., 2021; Laaksonen et al., 2023; Perhoniemi et al., 2023). The occupational trajectories before and after retirement in the case of City of Helsinki have been studied in Lallukka et al. (2023). See also Finnish Institute of Occupational Health for work ability forecast and work loading factors, and Laaksonen et al. (2018) and Shiri et al. (2021) studies on working days lost due to SAs or DP. An outline of the sickness benefit systems across Europe can be found in Spasova et al. (2016).

In this study we focus on early predictors of permanent disability. Using logistic regression model as an analytical tool we present results of a risk analysis depicting the occupation level risk for filing DP application in the near future. We also propose an indicator for the employers and pension providers, which utilizes up-to-date register data in revealing the critical duration of SA spells. The analysis is based on unique research data, which includes the necessary information about an employee's working life as well as his history of SA benefits and pension applications.

Keva is the sole Finnish public sector pension provider, with central and local government employers as customers. There are currently 1,940 employer customers of different sizes. Promoting digital services for employers with different HR resources, using a customer-centric approach is an essential task for Keva. Indeed, there is established literature on how design thinking and agile methodologies can lead to digitalization of public services and thus better government (e.g., World Bank, 2022; Welby et al., 2022). See also ISSA (2019).

Digital technologies already provide new possibilities for pension providers and employers. In Finland, there is ample and promising experience in using ICT in advancing the pension application processing (e.g., information gathering, stratifying pension applications, automatic decision making), thus shortening the application processing time. This development likely improves the trust in the pension system (e.g., Hyde et al., 2007; Vickerstaff et al., 2012). The overall possibilities of ICT are beyond the scope of this study, but we intend to deepen the understanding of how employers might benefit from statistical modeling and ICT in detecting the early indicators of disability risks. Up-to-date information is crucial in proactive management and is thus highly valuable for HR executives and occupational health care, as there are now web-based tools available for employers that assist in deepening their understanding of disability risk.

This study has three aims. One, we introduce a statistical model for DP retirement, based on simple yet comprehensive and up-to-date data. Two, we utilize the model in describing the factors affecting the risk of permanent disability and introduce an early indicator of diminishing work ability based on the SA days. Three, we show tentative results of the costs of SAs for one major local government employer, outlining the potential for developing digital tools for risk management.

**2. Disability pension schemes for the Finnish public sector employees**

The statutory pension system in Finland includes several pensions and benefits for those with weakened work ability (see SSA and ISSA, 2018). There are, in practice, two kinds of pensions: fixed term pensions (cash rehabilitation benefit and partial rehabilitation allowance) and those paid until further notice (disability pension and partial disability pension). The cash rehabilitation benefit or partial rehabilitation allowance is granted if it is estimated that the ability to work can be restored through treatment or rehabilitation. There is also a years-of-service pension available (since 2017) for those with permanently diminished work ability, but one is not yet eligible for DP.

The system rules indicate that an employee may be granted the abovementioned DPs and benefits if one has not reached the retirement age and if the ability to work has been reduced for at least one year because of an illness, an injury, or a disability. The rules also indicate that when drawing partial disability pension, one can earn in accordance with the personal salary earnings limit stated in the partial disability pension or partial disability cash rehabilitation benefit payment decision by the pension provider. Since January 2023 one can earn at least EUR 922/month (about 34 per cent of public sector median wage [see Finnish Centre for Pensions, 2021]). The years-of-service pension is granted for those older than 62 years old and one has worked for a long time (38 years or more) in strenuous and wearing work.

Beyond DPs there is also a benefit called occupational rehabilitation available for the working age employees with illness or injury. The benefit can be granted for those who have been employed over recent years but are at risk of having to be on DP in the next few years. Occupational rehabilitation includes several forms, for example, work trials, job coaching, education, apprenticeship training and career counselling.

There are also schemes and rules in place in the universal basic pension system to assist those with severely diminished work ability. The schemes include, for example, sickness allowance, rehabilitation allowance, rehabilitation, and disability pensions (rehabilitation subsidy or disability pension). The pensions can be granted after one has received sickness allowance for about one year. There are specific rules in the universal basic pension system for the DPs and benefits. As a prerequisite for DPs and benefits in statutory pension systems, a medical statement is required. See also an outline of the Finnish statutory pension system in Lähderanta et al. (2022) and OECD (2023).

From earlier studies we can observe that there are some differences in the disability retirement between public sector and private sector in Finland. The differences are mainly driven by occupational structure and gender along with age-structure and employer practices. Permanent disability pension is the most frequent DP in both sectors, but in the public sector it is more common to utilize fixed term pensions (see Polvinen, 2021; Polvinen and Laaksonen, 2023).

From Figure 1 we can see that the number of disability pension recipients has decreased notably (32 per cent) since 2017, which can be considered a positive trend. There is also a decrease in partial disability pensions (8 per cent) The number of partial disability pension retirees has increased by two per cent. In fact, the incidence rate of DPs in overall statutory earnings-related pension system has halved since 2002 (see Finnish Centre for Pensions, 2024a).

We can summarize that full disability pensions are still the major labor market outcome for those employees with persistent challenges in their work ability. It would be a positive development if the use of permanent disability could be hindered further and employees with health problems could maintain labor market attachment via fixed-term pensions and benefits. Furthermore, the partial disability pensions allow the interleaving of employment and pensions, which maintains decent labor market attachment. According to recent Finnish study the main reasons behind the decreasing trends in the DP recipients are on one hand decreasing amounts of new retirees and on the other hand the fact that DPs eventually switch to old-age pensions (OA). Mortality has no significant role in these trends (see Laaksonen and Rantala, 2023).

In the stock of DPs, the main causes of disability have been mental and behavioral disorders and diseases of the musculoskeletal system and connective tissue (see ICD-10, Finnish Centre for Pensions, 2024b).

Figure 1. Recipients of public sector statutory earnings-related disability pensions, persons

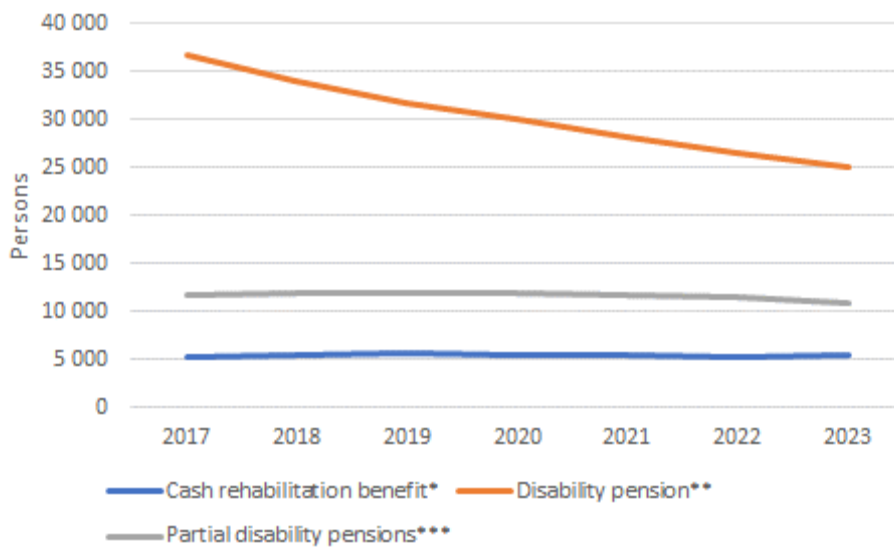

Notes: *=Fixed term, **=Paid until further notice, ***=Fixed term or paid until further notice.
Source: Keva's administrative registers, Authors.

**3. Research data, measurements and methods**

The initial data was drawn from the administrative registers of Finnish Public Sector Pension Provider Keva. Additional data on SA spells was received from certain public sector employers, that is, from major cities and wellbeing services counties. In total there were 21 employers that provided material for this study. Note that the novelty of our study is on the SA spells, which cover spells of all lengths.

The data set for the statistical modeling consisted of 940,021 observations from the years 2016 to 2019, with a gender distribution of 24.1 per cent male and 75.9 per cent female. These observations encompassed 340,816 individual employees. The data has been combined from three major separate data sources: SA leaves reported by the employers, employment information for those employers, and any pension benefits paid to the persons involved.

*Measurements of outcome and predictors*

*Outcome*

The outcome variable 'DP retirement' gets the values (0=no, 1=yes), and it was set to be true (yes) if any DP payments were paid (granted pension) to the person at any time during the three calendar years following the year of the observation. In total 19,816 observations (a prevalence of 2.1 per cent in the data set) had true outcome variable.

*Predictors*

The prepared data set has one row per person per year and gives information about the different length SA spells attributed to the person during the observation year, with overlapping or consecutive periods that had no gap between them concatenated provided their initiation years were the same, the gender and age of the person, and whether the person had previously received DP payments, as well as information about the principal employer and the primary occupation of the person during the year in question. The data set included all employed personnel, regardless of whether they had any SA spells or not. Any observations about a person already eligible for an OA pension, as well as observations where the person was receiving DP during the end of the observation year had already been removed while pre-processing the data set.

The explaining variables include socioeconomic information (Gender, Age <31 yrs., Age 43–53 yrs., Age, Occupation risk class), SA spell-specific information (SA spells (0–4 days), SA spells (5–9 days), SA spells (10–14 days), SA spells (15–29 days), SA spells (30–44 days), SA spells (45–59 days), SA spells (60+ days), Years since 60+ SA spell, SA spell (60+ days), Number of 60+ SA days), pension history information (Eligible for OA pension, Previous DP, Years since SA) and some auxiliary information (Year, Employer SA duration distribution). In the final model we also utilize some interactions of the explaining variables. See also Appendix Table 1 for more information about the content of all predictors.

As our final task is to provide tools for the employers, the required data is relatively limited and easy to update, yet it yields possibilities for making accurate predictions about the risk of DP.

*Method and study design*

As a statistical tool we use logistic regression to model the risk for drawing DP. Logistic regression is a technique which has been applied in many fields of science. In social and actuarial science, it has been applied to predictive tasks (e.g., Astari and Kismiantini, 2019; Wijnhoven et al., 2023; Altwicker-Hámori, 2023). In terms of content, similar modeling can be found in Betghe et al., (2021).

Our modeling was done by using R statistical software (R Core Team 2021; Wickham, 2016; Schauberger, 2023; Wickham et al., 2023a; Wickham et al., 2023b; Wickham et al., 2024; Sing et al., 2005) and SAS was used in data pre-processing and in summarizing the results.

The study design takes advantage of the data from the years 2016 to 2021, by pooling the years. In practice, one-year cross-sectional information of the predictors is used to predict the risk of drawing DP in the next three years (Figure 2). With the design we can study the in-sample probability for drawing DP (e.g., predictors from year 2016 and predictions for the years 2017 to 2019) and make out-of-sample predictions (e.g., predictors from year 2021 and predictions for the years 2022 to 2024).

Figure 2. Study design and data

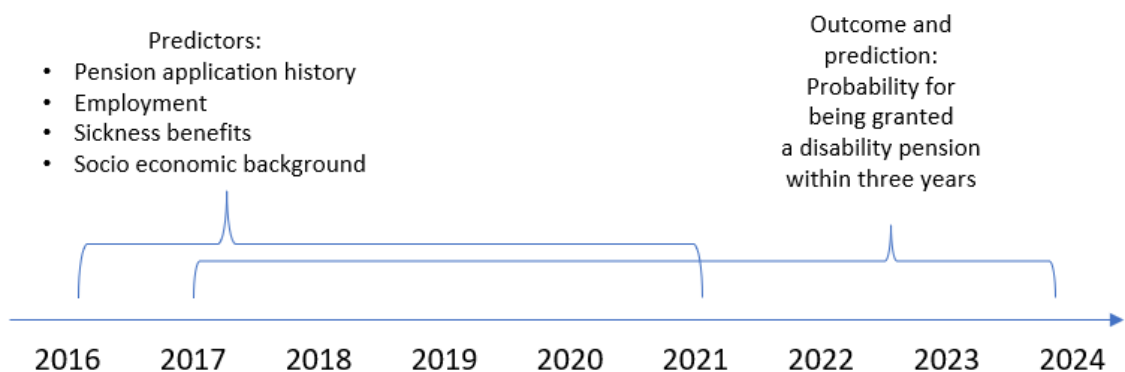

Source: Authors.

The data and study design allow us to study the following principal questions:

- Which public sector occupations are subject to increased disability pension risk?
- How can statistical modeling be utilized in providing tools for the employer?

**4. Modeling individual disability risk with logistic regression**

A set of model candidates of varying complexity were subjected to a backwards elimination process to find the most parsimonious model given the additional constraint of using only 30 or less predictors.

Since the sole purpose of the model was to predict the probability of the event, we also allowed for cases where the main effects were omitted in the predictor elimination process, and only interaction terms remained. In practice, a sample of 500,000 observations was marked as the teaching data set and a separate sample of 200,000 observations was used as testing and validation data set.

The model evaluation metrics included the AIC and McFadden Pseudo R-Squared relative rankings on the teaching data, and the relative AUC (Area Under the Curve) score ranking on the test data. Only models with 30 or less predictors after the stepwise predictor elimination process were considered for the final risk assessment model. The final selected logistic regression model utilized 'DP retirement' as the outcome and the abovementioned 20 explaining variables and their interactions (total number of predictors was 30). The model AUC score on the test sample was 0.84, which can be considered a decent value, considering the available explaining variables.

The explaining variable 'Occupation risk class' was constructed in order to take into account the differences in the ex-ante disability risk across occupations. All the public sector occupations were ordered into 10 groups according to the observed risk in the whole data set.

Finally, the selected model was re-trained for the final application with all the available observations (940,021) for whom the outcome variable was known. The final coefficients of the model were obtained from the full data set and are given in Table 1. The statistics indicate that most of the predictors have statistically significant value in the selected model.

Table 1. Model estimates and statistics

| Predictor | Estimate | Std. Error | z value |
| --- | --- | --- | --- |
| Intercept | -7.6300*** | 0.0772 | -98.80 |
| Year | -0.2029*** | 0.0370 | -5.49 |
| Eligible for OA pension | -1.7960*** | 0.0482 | -37.27 |
| SA spell (60+ days): 1 | 0.5493*** | 0.0412 | 13.34 |
| Previous DP: 1 | 1.6310*** | 0.0595 | 27.43 |
| Gender: female | 0.1154*** | 0.0231 | 4.99 |
| Age 43–53 yrs.: 1 | -0.6370*** | 0.0244 | -26.12 |
| Age | 0.0599*** | 0.0016 | 37.00 |
| Number of 60+ SA days | 0.0113*** | 0.0008 | 14.51 |
| SA spells (15–29 days) | 0.7728*** | 0.0417 | 18.52 |
| SA spells (30–44 days) | 1.0750*** | 0.1508 | 7.13 |
| SA spells (45–59 days) | 1.5470*** | 0.0800 | 19.33 |
| SA spells (60+ days) | 2.1730*** | 0.1330 | 16.34 |
| Year*Age | 0.0023*** | 0.0007 | 3.24 |
| SA spell (60+ days): 1*Years since 60+ SA spell | 0.0082 | 0.0186 | 0.44 |
| Previous DP*Years since SA | -0.0621*** | 0.0126 | -4.94 |
| Gender: female*Number of 60+ SA days | -0.0009 | 0.0009 | -1.09 |
| Gender: female*SA spells (15–29 days) | -0.1304*** | 0.0449 | -2.90 |
| Gender: female*SA spells (45–59 days) | -0.3267*** | 0.0878 | -3.72 |
| Gender: female*SA spells (60+ days) | -0.2385*** | 0.0742 | -3.21 |
| SA spells (0–4 days)*Age <31 yrs.: 0 | -0.0936*** | 0.0182 | -5.14 |
| SA spells (0–4 days)*Age <31 yrs.: 1 | 0.0039 | 0.0139 | 0.28 |
| SA spells (0–4 days)*Age 43–53 yrs.: 1 | 0.0049 | 0.0068 | 0.72 |
| Age 43–53 yrs.: 1*SA spells (30–44 days) | 0.1102** | 0.0572 | 1.93 |
| Age*SA spells (5–9 days) | 0.0048*** | 0.0002 | 26.57 |
| Age*SA spells (10–14 days) | 0.0089*** | 0.0003 | 28.85 |
| Age*SA spells (30–44 days) | -0.0040 | 0.0029 | -1.39 |
| Age*SA spells (60+ days) | -0.0114*** | 0.0022 | -5.13 |
| Age*Occupation risk class | 0.0021*** | 0.0001 | 24.25 |
| Age*SA spells (0–4 days)*Employer SA duration distribution | 0.0056*** | 0.0006 | 9.13 |

Note: Significance levels: ***=1%, **=5%, *=10%.
Source: Authors.

*Disability risk in major occupations*

Applying the regression model to individual-level data yields an estimate of the predicted risk of employees at a given point of time. This allows for building reports highlighting the variation and yearly change in the predicted risk across employee groups in a given employer. Furthermore, the model allows for tentatively evaluating the cost of disability risk.

For example, the estimated three-year DP risk of the employees at the end of year 2021 is presented across age and occupational groups below. The total three-year risk of a DP per employee being 1.82 per cent, but there is great variation across occupational groups as well as between age groups. As expected, the per-employee risk assessment increases with age. It is also noteworthy that in local government, many occupations are highly gendered, and age structures vary.

Table 2. Risk of disability pension by occupation, per cent

| Occupation* | Age 17 and older | Age 55 and older |
|---|---|---|
| Nurses | 1.96% | 6.16% |
| Practical nurses | 2.66% | 8.73% |
| Upper secondary school teachers | 0.92% | 3.24% |
| Childcare workers | 2.29% | 7.40% |
| Social service nurses and counselors | 1.63% | 5.76% |
| Experts in law, administration and economy | 1.28% | 3.30% |
| Construction, transportation and maintenance workers | 1.99% | 4.83% |
| Primary school teachers | 1.72% | 5.06% |
| Physicians | 0.86% | 3.07% |
| Office assistants and customer service workers | 3.04% | 6.70% |
| Elementary school teachers | 1.03% | 4.21% |
| Other nurses | 2.21% | 6.05% |
| Other workers | 0.93% | 3.16% |
| Hospital care assistants | 3.35% | 8.30% |
| Therapists in other healthcare | 2.06% | 6.08% |
| Social workers and other experts in social work | 1.52% | 3.93% |
| School assistants | 1.85% | 6.11% |
| Cleaning and kitchen workers | 3.05% | 7.13% |
| Firefighters and internal security officers | 1.28% | 5.08% |
| Experts in science and technology | 1.69% | 3.81% |
| Culture and art workers | 0.87% | 2.80% |
| Special teachers | 1.44% | 4.77% |
| Public health nurses | 1.62% | 4.83% |
| Other teaching experts | 1.50% | 3.48% |
| Library, museum and archives workers | 2.06% | 4.95% |
| Paramedics | 1.19% | 6.13% |
| *Total* | *1.82%* | *5.50%* |

Note: *ISCO-08-like classes.
Source: Authors.

Combined with evaluation of the cost of individual DPs, the individual-level risk ratings provide a tool for approximating the impact to the employer's "annual pension expenditure" should some or all the predicted risks realise. Also, the estimated effect of a change in some of the risk factors can be evaluated in concrete monetary terms.

*Critical duration of sickness absence days*

As employers need early indicators of diminishing work ability, Keva has proposed an indicator based on the duration of the SA days. The indicator captures the underlying measures of individual working life and the abovementioned risk factors. Based on statistical modeling of individual cases, it can be concluded that there is a certain sum of yearly SAs, which indicates a high risk for filing a DP application within three years. Here this yearly sum of days is called 'critical duration'.

In Finland, after 60 accrued SA days, there is a legal requirement for a physician's assessment of a person's work ability. However, considering the need for effectively recognizing those employees with the risk of work disability, most employers today have policies in place for early work ability intervention, where the frequency and amount of accrued SAs, combined with other information, are used as a guide for the HR department in timing their intervention. However, the guidelines concerning the actual critical values used in this process are varied and usually defined by expert opinion.

The research data allows us to model the critical duration of SA days. From Table 3 we can see the indicator in major public sector occupations. Note that the Table shows the results for those over 45 years of age with no previous DP application history and no long-term SA spells (60+ days) in a given year. As a second example of using risk assessment as a tool, the employees in the highest decile of estimated three-year disability risk are considered. In this example observations from years 2019–2021 are pooled to ensure a sufficient representation of all occupations. These highest-risk deciles over three years account for 57 per cent of the total combined three-year risk estimate, but only for 8 per cent of the total number of employees. Our task is to recognize the highest-risk group by examining the distribution of the accrued annual SA days.

As the number of SA days increases, the number of highest-risk employees in the remaining population decreases, but the share of these highest-risk cases in the remaining population increases. Assuming limited HR resources and the need to find the amount for the total of accrued annual SA days so that both the number and the share of highest-risk employees crossing this threshold are maximized, the results for this critical value for annual accrued SA days in occupational groups are presented below. The differences between occupational groups highlight the variance in the impact of the factors representing the number of SA days for the total risk in the prediction model.

Table 3. Critical duration of sickness absence days in major public sector occupations in 2021

| Occupation** | Share of total* | Critical duration, days |
|---|---|---|
| Secondary school teachers | 9% | 18 |
| Nurses | 8% | 20 |
| Practical nurses | 8% | 12 |
| Construction, transportation and maintenance workers | 7% | 16 |
| Childcare workers | 6% | 14 |
| Experts in law, administration and economy | 6% | 16 |
| Office assistants and customer service workers | 6% | 8 |
| Social service nurses and counselors | 5% | 17 |
| Primary school teachers | 4% | 20 |
| Physicians | 4% | 19 |
| Hospital care assistants | 4% | 8 |
| Cleaning and kitchen workers | 3% | 9 |
| Elementary school teachers | 3% | 20 |
| Social workers and other experts in social work | 3% | 16 |
| School assistants | 2% | 21 |
| Special teachers | 2% | 20 |
| *Total* | *100%* | *15* |

Notes: *Age 45 and older without long-term sickness or disability pension background, **ISCO-08-like classes.
Source: Authors.

Because DPs are very rare in the early life course, we focus on those over the age of 45. Also, in cases where there is a history of a DP or 60 annual SA days, the employer should always consider these employees being in high risk of disability. These cases are excluded from the analysis. With these limitations, and combined with other information about employees, the employee-specific guidelines for critical SA days can be used in timing early interventions. Other similar approaches may be used with different trajectories of SA days together with the critical duration.

From Table 3 we can see that among the participants in the study, the average critical duration is 15 days. We can also see that there is notable variation across the major occupations. The occupations with lower-than-average critical duration are interesting. In a range of social and healthcare occupations, the critical duration is low. Example of those occupations is office assistants and customer service workers (8 days). There are also some physically demanding occupations, for example, hospital care assistants (8 days), cleaning and kitchen workers (9 days), and practical nurses (12 days).

These demanding occupations with below average critical durations are no surprise since we already know from previous studies that they are both physically and mentally strenuous. The novelty is that the indicator can be observed daily in occupational health care.

Occupations with notably higher-than-average critical duration include nurses, pre-school teachers and elementary school teachers (20 days). We can assume that in these large occupational groups, the connection between SA and DP is weaker than in occupations with lower critical duration, that is, there is an inherent amount of SAs without it inferring the risk of DP.

### 5. Costs of sickness absences – A case study of major local government employer

The example case is one of the largest local government employers in Finland, employing wide range of professionals. The largest occupational groups are in healthcare, education, social services, and childcare, accompanied with administration and other support functions.

The proposed prediction model for DP risk, applied to the case employees in 2021, yields a snapshot of the distribution of the risk in different age and occupational groups, and allows comparison to other similar employers. The model yields an estimate of 556 new DPs during the next three years for the employer. The risk is most prominent in older age groups and occupations with the most pronounced stressors for disability. Occupational groups also differ in terms of age and gender. The risk estimates as well as some descriptives are presented in Table 4.

Table 4. Disability risk, critical duration and descriptive statistics.

| Occupation* | DP risk | Mean age, years | Mean SA, days | Share of women | Critical duration, days |
|---|---|---|---|---|---|
| Practical nurses | 1.7% | 41.4 | 13.8 | 85% | 17 |
| Upper secondary school teachers | 0.9% | 45.4 | 5.2 | 72% | 18 |
| Social service nurses and counselors | 1.3% | 42.4 | 11.3 | 80% | 23 |
| Childcare workers | 1.7% | 42.1 | 16.5 | 93% | 18 |
| Construction, transportation and maintenance workers | 1.6% | 45.3 | 9.7 | 26% | 22 |
| Experts in law, administration and economy | 1.2% | 46.3 | 4.8 | 68% | 20 |
| Primary school teachers | 1.2% | 42.4 | 14.0 | 95% | 26 |
| Nurses | 1.4% | 42.5 | 12.3 | 86% | 22 |
| Elementary school teachers | 0.6% | 41.3 | 9.2 | 79% | 20 |
| Office assistants and customer service workers | 2.1% | 47.6 | 10.3 | 87% | 14 |
| Social workers and other experts in social work | 1.1% | 44.2 | 6.5 | 88% | 23 |
| Cleaning and kitchen workers | 2.5% | 47.3 | 11.7 | 80% | 14 |
| Physicians | 0.8% | 41.3 | 5.1 | 74% | 21 |
| Other nurses | 2.1% | 45.8 | 12.5 | 87% | 17 |
| Public health nurses | 1.1% | 40.6 | 11.5 | 99% | 18 |
| Experts in science and technology | 1.3% | 47.2 | 5.3 | 42% | 19 |
| Special teachers | 1.1% | 45.1 | 10.4 | 85% | 23 |
| Therapists in other healthcare | 1.2% | 42.3 | 8.9 | 83% | 20 |
| School assistants | 1.4% | 42.5 | 14.2 | 80% | 25 |
| Firefighters and internal security officers | 1.6% | 41.6 | 15.0 | 9% | 22 |
| Library, museum and archives workers | 1.7% | 47.4 | 9.6 | 67% | 17 |
| Other teaching experts | 1.2% | 48.2 | 5.0 | 73% | 22 |
| Other workers | 1.2% | 44.5 | 4.6 | 46% | 21 |
| Culture and art workers | 1.0% | 44.1 | 4.8 | 54% | 32 |
| Hospital care assistants | 1.3% | 46.3 | 6.5 | 75% | 17 |
| Paramedics | 0.5% | 38.5 | 7.6 | 32% | 25 |
| *Total* | *1.4%* | *43.8* | *10.3* | *75%* | *19* |

Note: *ISCO-08-like classes.
Source: Authors.

In 2021, the employees faced almost 88,000 SA spells. About 84 per cent of these were 1–5 days long, 15 per cent lasted 6–30 days, and 1.1 per cent had a duration over 30 days. The total direct cost of sickness absences was estimated at 50 million euros in 2021, and the employer's disability pension payments totaled 20.3 million in 2021. If the number of SA spells was to decrease by 20 per cent, we can evaluate the effect on the estimated DP risk, and also get an approximation of how different costs would be affected when other risk factors are staying the same. In Table 5, the decrease in estimated DP risk as well as in direct cost of SAs and cost of pension payments for the employer are presented. These estimates highlight the cost of SAs and disability for the employer.

Table 5. Estimated effect of 20 per cent decrease in the number of sickness absence spells*

|  | 1–5 days | 6–30 days | Over 30 days |
| --- | --- | --- | --- |
| DP risk | -3.80% | -5.50% | -4.20% |
| Number of new DP retirees in three years | -21 | -30 | -23 |
| Direct cost of SA spells | -3.5 M€ | -2.5 M€ | -0.4 M€ |
| Employer pension payments | -1.3 M€ | -1.8 M€ | -1.4 M€ |

Notes: *The day intervals follow from Keva's already established digital services.
Source: Authors.

The effect of SA spells of different lengths on the estimated DP rate varies across employee and age groups. Usually, the effect of short spells (1–5 days) is smaller than that of longer spells, also younger employees' SA spells, in general, have the smallest effect on the DP risk. When comparing occupational groups, we can find groups for which there is a strong connection between SA spells and DP risk, for example, practical nurses under the age of 45 with SA spells over 6 days. On the other hand, the connection is weak, for example, for upper secondary school teachers, where only the SA spells over 30 days for employees over the age of 60 have a stronger connection to DP risk.

The critical duration for different occupations was also presented for the case in Table 4. For all employees at or over the age of 45, it is 19 days, somewhat higher than for the public sector total (cf. 14 days in Table 3). Comparing major occupations, the critical duration ranges from 14 days (cleaning and kitchen workers, and office assistants and customer service workers) to 32 days (culture and art workers). The occupations with lowest critical durations are also among the oldest on average. For some professions, the mean of annual SA days (not presented in table for ages 45 and older) is very near the critical duration, for example, for childcare workers, cleaning and kitchen workers, and practical nurses. These occupations should be given more attention by the employer, especially considering that the predicted risk is also among the highest in these professions.

## 6. Modeling and employer customer services

In Finland the providers of earnings-related statutory pensions constantly develop digital services for their employer customers especially regarding DPs. The key aim of the services is preventing permanent disability, for example by providing practical work ability services. Digital services include tools for evaluating the significance of SAs and their costs. As Keva is the sole pension provider of public sector pensions the aforementioned services are constantly developed in the institute.

The modeling framework presented in this paper yields a new step forward in establishing and delivering a concrete tool for analyzing the risk of DPs and their costs across public sector employers. Keva has already made further steps in taking advantage of ICT where the model estimates have been utilized in delivering several employer-specific indicators via Power BI visualization tools. The employers who have provided the necessary SA information find the new tool useful, and they are committed to providing the data also in the future.

## 7. Conclusions

The aim of this study is to give an overview of the modeling experiences in predicting future disability retirement in the context of statutory pension system. The modeling stems from Keva's statutory task to minimize new DPs and to introduce digital tools for the employer customers.

The research data and study design allow us to study SA spells of all lengths, which is a clear advantage in modeling disability risk. The modeling yields two major topics: estimates for occupation-specific DP retirement risk and critical duration of SA days, which indicates, in advance, highly increased risk for DP retirement in a given occupation.

According to the modeling results employers should pay attention to SAs in certain occupations, given socioeconomic information about the employees. At midlife the occupations with highest above the average risk were physical in nature. The critical duration indicate that 15 SA days is the limit over which the risk for DP retirement increases significantly. In some occupations, the critical duration is clearly smaller.

Generalizability of the modeling results is a central question. Keva already provides tools for the employers to summarize, for example, occupational health care costs. As the employers are of different sizes, the snapshot of the work ability implications on the finances varies. Large employers, for example, major municipalities and wellbeing services counties have up-to-date information of the costs. Small employers, on the other hand, have limited resources and capabilities to initiate new practices. We should note that the data used in developing the proposed digital tool does not include small employers' data and therefore the results must be used with caution. The already established tools have given employers useful and uniform information as well as support for practical situations in the workplace. The proposed tool has drawn attention and interest among the employers and beyond those who provided the necessary data.

As a general note, studies have indicated that employers can gain net benefits from their investments in work disability management, for example. by targeting management actions to the main work disability risks (e.g., Leino et al., 2023; Reiman et al., 2017). Large part of the possible problems in corporate wellness can, in practice, be solved by the workplace's own means, such as by improving strategic work ability management and anticipating the decrease of health among employees (e.g., Anttilainen et al., 2023).

**References**

Altwicker-Hámori, S. 2023. Labour market integration of disability insurance benefit applicants in Switzerland Evidence from linked survey and administrative data. ALTER – 17/1 (2023) – 69–86.

Anttilainen, J.; Pehkonen, I.; Savinainen, M.; Haukka, E. 2023. Social and health care top managers' perceptions and aims of strategic work ability management in the midst of change. Work. DOI:10.3233/WOR-230034

Astari, D.; Kismiantini. 2019. Analysis of Factors Affecting the Health Insurance Ownership with Binary Logistic Regression Model. J. Phys.: Conf. Ser. 1320 012011. DOI 10.1088/1742-6596/1320/1/012011

Bethge, M.; Spanier, K.; Streibelt, M. 2021. Using Administrative Data to Assess the Risk of Permanent Work Disability: A Cohort Study. J Occup Rehabil. 2021; 31(2): 376–382. doi: 10.1007/s10926-020-09926-7

Finnish Centre for Pensions. 2021. https://www.etk.fi/en/research-statistics-and-projections/statistics/insured-persons/

Finnish Centre for Pensions. 2024a. https://tilastot.etk.fi/pxweb/en/ETK/ETK__120tyoelakkeensaajat__55tyokyvyttomyyden_alkavuus/?tablelist=true

Finnish Centre for Pensions. 2024b. https://tilastot.etk.fi/pxweb/en/ETK/ETK__120tyoelakkeensaajat__20tyoelakkeensaajien_lkm/elsa_t10_tk_diag.px/

Finnish Institute of Occupational Health. https://worklifedata.fi/en/data/work-ability-forecast

Hyde, M.; Dixon, J.; Drover, G. 2007. Assessing the Capacity of Pension Institutions to Build and Sustain Trust: A Multidimensional Conceptual Framework. Journal of Social Policy, Volume 36, Issue 3, July 2007, pp. 457–475. DOI: https://doi.org/10.1017/S0047279407001043

ISSA. 2019. ISSA Guidelines on service quality. Geneva, International Social Security Association.

Keva. https://www.keva.fi/en/

Laaksonen, M.; Rantala, J.; Järnefelt, N.; Kannisto, J. 2018. Educational differences in years of working life lost due to disability retirement. Eur J Public Health 28(2):264–268. https://doi.org/10.1093/eurpub/ckx221

Laaksonen, M.; Rantala, J. 2023. Mistä työkyvyttömyyseläkkeellä olevien määrän väheneminen johtuu?. Yhteiskuntapolitiikka 88 (2023):4. https://urn.fi/URN:NBN:fi-fe20230913125212

Laaksonen, M.; Blomgren, J.; Rinne, H.; Perhoniemi, R. 2023. Impact of a Finnish reform adding new sickness absence checkpoints on rehabilitation and labor market outcomes: an interrupted time series analysis. Scand J Work Environ Health. https://doi.org/10.5271/sjweh.4122

Lallukka, T.; Lahelma, E.; Pietiläinen, O.; Kuivalainen, S.; Laaksonen, M.; Rahkonen, O.; Lahti, J. 2023. Trajectories in physical functioning by occupational class among retiring women: the significance of type of retirement and social and health-related factors. Journal of Epidemiology & Community Health, vol. 77, no. 6, pp. 362–368. https://doi.org/10.1136/jech-2022-219963

Leino, T.; Turunen, J. K. A.; Pehkonen, I.; Juvonen-Posti, P. 2023. Important collaborative conditions for successful economic outcomes of work disability management: A mixed methods multiple case study. Work 2023;74(2):685–697. doi: 10.3233/WOR-210026

Leino-Arjas, P.; Seitsamo, J.; Nygård, C.H.; Prakash, K.C.; Neupane, S. 2021. Process of Work Disability: From Determinants of Sickness Absence Trajectories to Disability Retirement in A Long-Term Follow-Up of Municipal Employees. International Journal of Environmental Research and Public Health, 18(5), (2021), 2614. doi.org/10.3390/ijerph18052614

Lähderanta, T.; Salonen, J.; Möttönen, J.; Sillanpää, M. J. 2022. Modelling old-age retirement: An adaptive multi-outcome LAD-lasso regression approach. International Social Security Review, Vol. 75, 1/2022. https://doi.org/10.1111/issr.12287

Mailes, G.; Carrasco, M.; Arcuri A. 2021. The global trust imperative. [S.l.]. Boston Consulting Group & Salesforce.

OECD. 2022. Disability, Work and Inclusion: Mainstreaming in All Policies and Practices. OECD Publishing, Paris, https://doi.org/10.1787/1eaa5e9c-en.

OECD. 2023. Pensions at a Glance 2023: OECD and G20 Indicators.

Perhoniemi, R.; Blomgren, J.; Laaksonen, M. 2023. Identifying labour market pathways after a 30-day-long sickness absence -a three-year sequence analysis study in Finland. BMC Public Health. 2023 Jun 7;23(1):1102. doi: 10.1186/s12889-023-15895-2.

Polvinen, A. 2021. Työkyvyttömyyseläkkeelle siirtymisen erot kunta-alan ja yksityisen sektorin palkansaajilla. Kuntoutus, 44(1), 10–23. https://doi.org/10.37451/kuntoutus.103338

Polvinen, A.; Laaksonen, M. 2023. Contribution of age, gender and occupational group to the higher risk of disability retirement among Finnish public sector employees. Scand J Public Health. 2023 Feb 22:14034948231153913. doi: 10.1177/14034948231153913

R Core Team 2021. R: A language and environment for statistical computing. R Foundation for Statistical Computing, Vienna, Austria. URL https://www.R-project.org/

Reiman, A.; Ahonen, G.; Juvonen-Posti, P.; Heusala, T.; Takala, EP.; Joensuu, M. 2017. Economic impacts of workplace disability management in a public enterprise, International Journal of Public Sector Performance Management (2017);3: (3):297.

Schauberger, P. 2023. Package 'openxlsx'. https://cran.r-project.org/web/packages/openxlsx/openxlsx.pdf

Shiri, R.; Hiilamo, A.; Rahkonen, O.; Robroek, S. J. W.; Pietiläinen, O.; Lallukka, T. 2021. Predictors of working days lost due to sickness absence and disability pension. International Archives of Occupational and Environmental Health, vol. 94, no. 5, pp. 843–854. https://doi.org/10.1007/s00420-020-01630-6

Sing, T.; Sander, O.; Beerenwinkel, N.; Lengauer, T. 2005. ROCR: visualizing classifier performance in R. Bioinformatics 21(20):3940-1.

Spasova, S.; Bouget, D.; Vanhercke, B. 2016. Sick pay and sickness benefit schemes in the European Union, Background report for the Social Protection Committee's In-depth Review on sickness benefits (17 October 2016). European Social Policy Network (ESPN), Brussels, European Commission.

SSA; ISSA. 2018. Social security programs throughout the world: Europe. Woodlawn, MD, Social Security Administration.

Welby, B.; Tan, E. 2022. Designing and delivering public services in the digital age, Going Digital Toolkit Note, No. 22, https://goingdigital.oecd.org/data/notes/No22_ToolkitNote_DigitalGovernment.pdf.

Vickerstaff, S.; Macvarish, J.; Taylor-Gooby, P.; Loretto, W.; Harrison, T. 2012. Trust and confidence in pensions: A literature review. UK, Department for Work and Pensions. Working Paper no. 108.

Wickham, H. 2016. ggplot2: Elegant Graphics for Data Analysis. Springer.

Wickham, H.; Miller, E.; Smith, D. 2023a. haven: Import and Export 'SPSS', 'Stata' and 'SAS' Files. https://haven.tidyverse.org/index.html

Wickham, H.; François, R.; Henry, L.; Müller, K.; Vaughan, D. 2023b. dplyr: A Grammar of Data Manipulation. https://dplyr.tidyverse.org/index.html

Wickham, H.; Vaughan, D.; Girlich, M. 2024. tidyr: Tidy Messy Data. https://tidyr.tidyverse.org/index.html

Wijnhoven, M.; Dusseldorp, E.; Guiaux, M.; Havinga, H. 2023. The Work Profiler: Revision and maintenance of a profiling tool for the recently unemployed in the Netherlands. International Social Security Review. Volume76, Issue 2, April/June 2023 Pages 109–134. https://doi.org/10.1111/issr.12327

World Bank. 2022. Service Upgrade – The GovTech approach to citizen centred services. Washington, DC, World Bank Group.

**Appendix**

Table 1. Research data contents

| Explaining variable | Description |
|---|---|
| Occupation risk class | Relative risk class of the person's occupation on a scale of 1–10. In detail, the occupations are based on ISCO-08 classification, which is fine-tuned for Finnish public sector employees |
| Eligible for OA pension | Value indicating if the person is soon eligible for old-age pension (0=not eligible within the following three subsequent calendar years, 1/3=eligible in the third subsequent calendar year, 2/3=eligible in the second subsequent calendar year, 1=eligible in next calendar year) |
| Age <31 yrs. | Value indicating whether the integer part of the person's age at the end of the year of the observation is less than or equal to 30 |
| Age 43–53 yrs. | Value indicating whether the integer part of the person's age at the end of the year of the observation is greater than or equal to 43 and less than or equal to 53 |
| Age | Integer part of the person's age at the end of the year of the observation |
| SA spells (0–4 days) | Number of distinct continuous SA spells (concatenated if consecutive and started on the year of the observation) of length of at most 4 days |
| SA spells (5–9 days) | Number of distinct continuous SA spells (concatenated if consecutive and started on the same calendar year) of length of at least 5 days and at most 9 days |
| SA spells (10–14 days) | Number of distinct continuous SA spells (concatenated if consecutive and started on the same calendar year) of length of at least 10 days and at most 14 days |
| SA spells (15–29 days) | Number of distinct continuous SA spells (concatenated if consecutive and started on the same calendar year) of length of at least 15 days and at most 29 days |
| SA spells (30–44 days) | Number of distinct continuous SA spells (concatenated if consecutive and started on the same calendar year) of length of at least 30 days and at most 44 days |
| SA spells (45–59 days) | Number of distinct continuous SA spells (concatenated if consecutive and started on the same calendar year) of length of at least 45 days and at most 59 days |
| SA spells (60+ days) | Number of distinct continuous SA spells (concatenated if consecutive and started on the same calendar year) of length of at least 60 days |
| SA spell (60+ days) | Whether the person is known to have been on a continuous SA spell of at least 60 days during any of the preceding calendar years (0/1) |
| Previous DP | Whether the person has received disability pension at any time before the end of the year (0/1) |
| Number of 60+ SA days | How many days beyond the 60th day the longest continuous SA spell that started on the year of the observation lasted |
| Gender | Gender of the person (M=male, N=female) |
| Years since 60+ SA spell | How many years since the last preceding calendar year that the person was known to have been on a continuous SA spell for at least 60 days (consecutive SA spells were concatenated if they started on the same calendar year) |
| Employer SA duration distribution | Gini coefficient for the distribution of sick leave lengths reported by the persons principal employer during the year of observation |
| Years since SA | How many calendar years has there been from the last disability pension payments (0=last payment during the year of observation, 1=last payment during the preceding calendar year,…,10=last payment on the 10th preceding calendar year or earlier) |
| Year | Calendar year counter for the observation, 1=2016, 2=2017,… |